\documentclass[aps,prd,twocolumn,showpacs,nofootinbib,longbibliography,10pt]{revtex4-1}%
\usepackage[colorlinks,citecolor=DarkGreen,linkcolor=DarkRed,urlcolor=DarkBlue]{hyperref}
\usepackage{amsmath}
\usepackage{graphicx}
\usepackage{slashed}
\usepackage[svgnames]{xcolor}
\newcommand{\bs}{\boldsymbol}
\begin{document}

\title{Nucleon axial-vector coupling constant in magnetar environments}

\author{C. A. Dominguez$^1$}
\author{Marcelo Loewe$^{2,3,1}$}
\author{Cristian Villavicencio$^{4}$}
\email[E-mail: ]{cvillavicencio@ubiobio.cl}
\author{R. Zamora$^{5,6}$}

\affiliation{$^1$Centre for Theoretical \& Mathematical Physics, and Department of Physics, University of Cape Town, Rondebosch 7700, South Africa;}
\affiliation{$^2$Facultad de Ingenier\'ia, Arquitectura y Diseño, Universidad San Sebasti\'an, Santiago, Chile;}
\affiliation{$^3$Centro Cient\'ifico Tecnol\'ogico de Valparaíso-CCTVAL,
Universidad T\'ecnica Federico Santa Mar\'ia, Casilla 110-V, Valpara\'iso, Chile;}
\affiliation{$^4$Centro de Ciencias Exactas \& Departamento de Ciencias B\'asicas, Facultad de Ciencias, Universidad del B\'io-B\'io,
Casilla 447, Chill\'an, Chile;}
\affiliation{$^5$Instituto de Ciencias B\'asicas, Universidad Diego Portales, Casilla 298-V, Santiago, Chile;}
\affiliation{$^6$Centro de Investigaci\'on y Desarrollo en Ciencias Aeroespaciales (CIDCA), Academia Politécnica Aeronáutica, Fuerza A\'erea de Chile, Casilla 8020744, Santiago, Chile.}

\begin{abstract}
    The nucleon axial-vector coupling constant $g_A$ is studied in the presence of an external magnetic field, and in dense nuclear environments, to emulate nuclear matter in magnetars. 
    For this purpose we use QCD finite energy sum rules for two-current and three-current correlators, the former involving nucleon-nucleon correlators and the latter involving proton-axial-neutron currents.
    As a result, the axial-vector coupling constant decreases both with baryon density as well as with magnetic field. 
    The axial-vector coupling evaluated with baryon density near the nuclear density $\rho_0$ leads to $g_A^*\approx 0.92$. 
    In the presence of magnetic fields  $g_A$ decreases in general, but $g_A^*$ does not show significant changes.    
\end{abstract}

\maketitle

The nuclear axial-vector charge or axial-vector coupling constant $g_A$ of the nucleon plays an important role in compact star phenomenology. 
In particular, $g_A$ is related to nuclear beta decay processes, being the neutron decay width $\sim 1+3g_A^2$ \cite{Mund:2012fq,Czarnecki:2018okw}. 
Also, in neutrons stars electronic capture produces the inverse process which altogether is known as the Urca process. 
These decay and capture processes keep the star in chemical equilibrium with the constant loss of neutrinos.  
The neutrino emissivity also involves $g_A$ in the same way as in the neutron decay width  \cite{Yakovlev:2000jp}. 
Therefore, it is relevant to know how $g_A$ is modified in a compact star environment, in particular for a high baryon density.
It is accepted that $g_A$ is quenched in nuclear medium \cite{Rho:1974cx,Wilkinson:1974huj,Brown:1978zz,Park:1997vv,Carter:2001kw,Lu:2001mf,Suhonen:2017krv,Bass:2020bkl,Rho:2021zwm,Rho:2023vow}, and several studies have concluded that $g_A^*\equiv g_A(\rho_0)\approx 1$, where $\rho_0=0.16\text{ fm}^{-3} = 1.22\times 10^{-3}\text{ GeV}^{3}$ is the nuclear baryon density.

In magnetars it is expected that the magnetic field increases in the interior of the star with values that can reach $\sim 10^{19}$\,G in the core, which is equivalent to $eB \sim 3.4\, m_\pi^2$ \cite{Kaspi:2017fwg}.
Hence, such an intense magnetic field can modify considerably the hadronic parameters.
At zero baryonic density, the magnetic field decreases the axial-vector coupling constant \cite{Villavicencio:2022gbr,Braghin:2018drl}.
The influence of the magnetic field on the axial-vector coupling behavior in dense baryonic matter is the purpose of this work. We use finite energy sum rules (FESR) involving the three-current correlator (proton, axial-vector, neutron) in order to determine the magnetic and density evolution of the axial-vector coupling constant. 
In addition, we also determine the two-nucleon (proton-proton and neutron-neutron) current correlators in the presence of an external magnetic field, and in a dense baryonic medium,  to find the medium dependent evolution of the nucleon-current coupling and the continuum hadronic thresholds.

\section{FESR}
\label{sec.FESR}

Among the different types of sum rules  in the literature, FESR establish a clear criterion for the separation between the hadronic and the QCD sectors. 
Considering a form factor $\Pi(s)$ generated by current correlation, with $s=q^2$, 
the FESR are obtained by integrating the form factor along the well-known {\it pac-man} contour, i.e. a circle in the complex $s$ plane, with a cut on the positive real axis. 
Hadronic resonances enter the positive real axis, while the QCD sector is  on the circle. From Cauchy's theorem,  quark-hadron duality relates both sectors, i.e.
\begin{equation}
    \int_0^{s_0}\frac{ds}{\pi} s^N \text{Im}\Pi^\mathrm{\tiny had}(s)
   =-\oint_{s_0}\frac{ds}{2\pi i}s^N\Pi^\mathrm{\tiny QCD}(s) \,,     \end{equation}
where $s_0$ is the radius of the circle. 
The factor $s^N$ is an analytic kernel.

In order to relate these two sectors, nonperturbative effects must be included. 
They are  parametrized in terms of the operator product expansion (OPE) leading to
\begin{equation}
    \Pi(s) = C_\mathrm{\tiny pert}(s) + \sum_{n>0}C_n(s)\frac{\langle O_n\rangle}{s^n},
\end{equation}
where the first term corresponds to the perturbative sector, including radiative corrections from perturbation theory.
The OPE is a parametrization of nonperturbative effects in terms of background gluon and quark fields. 
For this purpose the radius of this circle, $s_0$, should be large enough. When radiative corrections, and medium effects are absent, the analytic kernel depending on $N$ eliminates most condensate contributions. 
This feature is quite convenient in order to identify the relevant condensates. 
When medium effects are present, usually Lorentz symmetry is broken, generating contributions from new condensates.
It is then necessary  to choose a frame to perform the integral as now $\Pi(p^2) \to \Pi(p_0,\bs{p})$.
The usual choice is to set $\bs{p}=0$ and $s=p_0^2$, by separating even and odd contributions of $\Pi(p_0)=\Pi^e(s)+p_0\Pi^o(s)$. 
\footnote{Alternatively, one could perform a contour integration in the complex-$p_0$ plane \cite{Furnstahl:1992pi}}

Here, we will use the usual contour for the nucleon-nucleon correlator, but calculate  the nucleon\,-\,axial-vector\,-\,nucleon correlator. 
In this case, a double FESR must be used \cite{Villavicencio:2022gbr,Dominguez:1999ka}
\begin{multline}
 \int_0^{s_p}\frac{ds'}{\pi} \,\mathrm{Im}_{s'}\!\!\int_0^{s_n}\frac{ds}{\pi}\,\mathrm{Im}_s\Pi^\text{\tiny had} (s,s',t)\\
 =\oint_{s_p}\frac{ds'}{2\pi i}\oint_{s_n}\frac{ds}{2\pi i}\,\Pi^\text{\tiny QCD}(s,s',t),
 \label{eq.FESR_had=QCD}
\end{multline}
where 
\begin{equation}
 \mathrm{Im}_s f(s)\equiv\lim_{\epsilon\to 0}\mathrm{Im}f(s+i\epsilon).
\end{equation}

The FESR in vacuum involve condensates of different dimensions. However, for in-medium scenarios this correspondence is no longer valid . This is due to the appearance of new condensates whose corresponding Wilson coefficients are e.g. powers of the form $\sim\ln(-s)/s^{n}$. Nevertheless, once FESR are implemented, these higher dimension operators are suppressed by powers of $s_0^{-n}$. 
If $s_0$ decreases, its is expected that the condensates  behave similarly, as known e.g. at finite temperature where condensates {\it melt}. 
In summary, vacuum FESR determine up to whatever dimension in the OPE is needed in applications.\\
Another feature of FESR is the fact that the contour integral can be exchanged with the momentum loop integral of the form factors. 
This is extremely useful because it is not necessary to determine the correlator up to its final form. 
In addition, it does not impose the condition of expansion in powers of $1/s$,  as usually done in OPE. This allows us to achieve low $s_0$ values.
For detailed information see e.g. \cite{Dominguez:2018njv,Villavicencio:2020fcz,Dominguez:2020sdf}.

\section{Axial-vector coupling in vacuum}
\label{sec.gA-vac}

Based on \cite{Villavicencio:2022gbr}, we discuss a  determination  of the axial-vector coupling of the nucleon in the framework of QCD sum rules. 
The main object of interest is the following currents correlator
\begin{equation}
 \Pi_\mu(x,y,z)=-\langle 0|\,{\cal T}\,\eta_p(x)A_\mu(y)\,\bar\eta_n(z)\,|0\rangle, 
\end{equation}
where in the QCD sector $\eta_N$ is Ioffe's nucleon interpolating current \cite{Ioffe:1981kw}
\begin{align}
 \eta_p(x) &=\epsilon^{abc}\left[u^a(x)^T\,C\gamma^\mu \,u^b(x)\right]\gamma_\mu\gamma_5 \,d^c(x),\\
 \bar\eta_n(z) &= \epsilon^{abc}\left[\bar d^b(z)\,\gamma^\mu C\,\bar d^a(z)^T\right]\bar u^c(z)\,\gamma_\mu\gamma_5 ,
\end{align}
where $C=i\gamma_0\gamma_2$ is the charge conjugation operator.

In the hadronic sector $\eta_p(x)$ is defined as
\begin{align}
 \langle 0|\,\eta_p(x)\,|p',s'\rangle &= \lambda_p\, u_p^{s'}(p')\,e^{-ip'\cdot x},\label{eq.eta_p} \\
 \langle p,s|\,\bar\eta_n(z)\, |0\rangle &= \lambda_n\, \bar u_n^s(p)\, e^{ip\cdot z},
\end{align}
where $\lambda_p$ and $\lambda_n$ are the phenomenological current-proton and current-neutron couplings, respectively.
Alternatively, we could introduce nucleonic fields whose matrix components are the ones described in the above equations, i.e.
\begin{align}
    \eta_N(x)=\lambda_N\Psi(x),
\end{align}
where $\Psi_N$ corresponds to the nucleonic fields, and $A_\mu$  to the positively charged axial-vector current.

In the QCD sector $A_\mu(y)$ is defined as
\begin{align}
A_\mu(y) = \bar d(y)\,\gamma_\mu\gamma_5\, u(y).
\end{align}

In the hadronic sector this is defined as
\begin{align}
 \langle p',s'|A_\mu(y)|p,s\rangle = \bar u_p^{s'}(p')\,T_\mu(q)\,u_n^s(p) \,e^{iq\cdot y},
\end{align}
with $q=p'-p$.

The function $T$ is the most general function in terms of the Clifford basis, compatible with the axial-vector sector and their relevant transformations, written as
\begin{equation}
 T_\mu(q) = 
 G_A(t)\gamma_\mu\gamma_5 +G_P(t)\gamma_5 \frac{q_\mu}{2  m_N}
 +G_T(t)\sigma_{\mu\nu}\gamma_5  \frac{q_\nu}{2 m_N},
 \label{T_mu}
\end{equation}
 with $t=q^2$  and  $ m_N$ the vacuum nucleon mass. Alternatively, the axial-vector current might  also be expressed in terms of nucleonic fields as
\begin{align}
A_\mu(y) = \int d^4\xi\, \bar\psi_p(\xi)\tilde T_\mu(\xi-y)\psi_n(\xi) ,
\end{align}
where $\tilde T_\mu(x)$ is the inverse Fourier transformation of $T_\mu(q)$ in configuration space.
 
The function  $G_A$ provides the definition of the axial-vector according to
\begin{equation}
g_A\equiv G_A(0).
\end{equation}

Fourier transforming the three-current correlator in momentum space leads to
\begin{equation}
 \Pi_\mu(p,p')=\int d^4y\, d^4z\, e^{-i(q\cdot y+p\cdot z)}\,\Pi_\mu(0,y,z).\label{Eq.Pi-momentum}
\end{equation}

The correlator in the hadronic sector can be expressed as usual by inserting a complete set of intermediate nucleon states , or by expressing the currents in terms of nucleon fields, leading to
\begin{equation}
 \Pi^\text{\tiny had}_\mu(p,p')=\lambda_n\lambda_p\frac{ (\slashed{p}+m_n)T_\mu(q)(\slashed{p}'+m_p)}{(p^2-m_n^2)(p'^2-m_p^2)}.
\end{equation}

Using an appropriate procedure one can isolate the $G_A$ contribution from the other form factors, e.g.
\begin{equation}
 \mathrm{tr}\,[\Pi_\mu(p,p')\,\gamma_\nu]=-4i\epsilon_{\mu\nu\alpha\beta}p^\alpha p'^\beta \Pi(s,s',t),
\end{equation}
with $s=p^2$, $s'=p'^2$, and  $\Pi$ in the hadronic sector being
\begin{equation}
 \Pi^\text{\tiny had}(s,s',t)=\lambda_n\lambda_p\frac{G_A(t)+G_T(t)(m_n-m_p)/ m_N}{(s-m_n^2)(s'-m_p^2)}.
 \label{Pi_had}
\end{equation}

In vacuum, proton and neutron masses are almost identical due to $SU(2)$ isospin symmetry.
In principle it is possible to completely isolate $G_A$ if the masses are different. This procedure could be more cumbersome than
the one introduced here. Even in the case of different masses, the contribution to $G_A$ proportional to $(m_n-m_p)/m_N$  would be small, provided usual isospin breaking is generated by external magnetic fields. 
If isospin density effects are considered, the mass difference  could be considerable.\\

Following the same procedure in the QCD sector, the two-loop leading order perturbative contribution in the frame $t=0$ becomes
\begin{multline}
 \Pi^\text{\tiny pQCD}(s,s',0)= 
 \frac{s^2\ln(-s/\mu^2)-s'^2\ln(-s'/\mu^2)}{(2\pi)^4\,(s'-s)}\\
 +\text{regular terms},
 \label{PI-pQCD}
\end{multline}
where $\mu$ is the $\overline{\text{MS}}$ scale. Regular terms means terms without discontinuities on the real axes, or singularities which would vanish in a FESR framework.\\ 
Using the double FESR described previously one obtains
\begin{multline}
 g_A\lambda_n\lambda_p
=\frac{1}{48\pi^4}\left[s_n^3\,\theta(s_p-s_n)
+s_p^3\,\theta(s_n-s_p)\right],
\label{eq.gA(p,n)}
\end{multline} 
where $s_N$ are the nucleon current hadronic thresholds, which must be bigger than $m_N^2$.
In vacuum, both nucleon thresholds are the same, therefore
\begin{equation}
g_A=\frac{s_0^3}{48\pi^4\lambda_N^2}
\end{equation}
which is the leading contribution, as the next correction, $\sim \langle G^2\rangle /s_0^2$, is negligible.\\
The information needed for completion is the values of the threshold and the nucleon coupling,
which will be determined using the nucleon-nucleon correlator \cite{Ioffe:1981kw,Reinders:1984sr,Sadovnikova:2005ye,Nasrallah:2013ywh}
\begin{equation}
\Pi_N(x)=\langle 0|\,{\cal T}\,\eta_N(x)\,\bar\eta_N(0)\,|0\rangle.
\end{equation}

From the FESR, the two Dirac structures become
\begin{align} 
\lambda_N^2 &= \frac{s_0^3}{192\pi^4}+\frac{s_0}{32\pi^2}\langle G^2\rangle+\frac{2}{3}\langle\bar qq\rangle^2
\label{eq.Nuclear_FESR_p}
\\
 \lambda_N^2m_N &=-\frac{s_0^2}{8\pi^2}\langle\bar qq\rangle+ \frac{1}{12}\langle G^2\rangle\langle\bar qq\rangle,
 \label{eq.Nuclear_FESR_s}
\end{align}
where the saturation approximation was invoked in operators of dimension d=6 and d=7.

Hence, the nucleon couplings and the hadronic thresholds are functions of the quark and gluon condensates.
In particular, as was pointed out in \cite{Villavicencio:2022gbr}, there is a small window for possible values of quark and gluon condensates  agreeing with the experimental values of the axial-vector coupling constant $g_A\approx 1.275$. We will use two sets of parameters that fit this experimental value.\\

When medium effects are taken into account, it is expected that the the axial-vector contribution to the form factor splits into different components, according to the symmetry breaking induced by external magnetic field and thermal or dense bath
\begin{equation}
    G_A\gamma_\mu\to G_A^{(0)}\gamma_0+G_A^{(\perp)}\gamma^\perp_\mu+G_A^{(3)}\gamma_3+\tilde G_A F_{\mu\nu}\gamma^\nu.
\end{equation}

We will limit ourselves to the leading contribution described by Eq.\,(\ref{eq.gA(p,n)}).
Hence, density and magnetic contributions will be present only through the in-medium nucleon-current couplings and nucleon hadronic thresholds.

\section{finite density}

Baryonic density effects in QCD sum rules for baryons were first considered in a series of pioneering papers  \cite{Cohen:1991js,Cohen:1991nk,Furnstahl:1992pi,Jin:1993up,Jin:1994bh}.
The basic idea is that density dependence enters only in the effective operators in the nonperturbative sector. 
The introduction of a dense medium will break Lorentz symmetry. Hence, the structure of correlators changes and new condensates appear. 
All  correlator structures can be separated into  {\it even}  and  {\it odd}   contributions in terms of $p_0$ as
\begin{align}
\Pi(p_0,\bs{p}^2) &= 
\Pi^e(p_0^2,\bs{p}^2)+p_0\Pi^o(p_0^2,\bs{p}^2).
\label{eq.even-odd}
\end{align}

The sum rules will involve the frame  where $\bs{p}=0$, defining $s=p_0^2$.
In the case of the nucleon-nucleon correlator the general structure is
\begin{align}
    \Pi_N = \Pi_s+\slashed{p}\Pi_p+\slashed{u}\Pi_u \,
\end{align}
where the subscript $s$ stands for scalar, the subscript  $p$ means proportional to $p$, and $u$ is the fourth velocity vector which in the system's rest frame is $u=(1,0,0,0)$.
Each of the above terms will be separated into even and odd contributions.\\

The hadronic sector is usually represented in terms of the finite density fermion self energy.
We write the nucleon correlator in the hadronic sector in a different way, which can easily be obtained in terms of the self energy
\begin{equation}
    \Pi_N = \frac{-\lambda_N^2}{\gamma_0 (p_0 +\Delta\mu)-v\,\bs{\gamma}\cdot \bs{p} -m_N},
    \label{eq.Pi_N-had}
\end{equation}
where $\Delta \mu $ is a correction to the baryon chemical potential, and $v$ is the fermion velocity.
The nucleon-current coupling and the nucleon mass are now density dependent, and in the absence of any isospin breaking factor, proton and neutron correlators will lead to the same results.

In the hadronic and  in the QCD sector, all possible structures were included. 
However, for our analysis we need only two of those terms in order to find $\lambda_N$, and also to avoid the presence of other condensates of dimension four, which are unknown, as well as higher dimensional condensates. 
In the case where we use Ioffe's interpolating nucleon current we have 
for proton up to dimension\,\mbox{}6 \cite{Cohen:1991js,Cohen:1991nk,Furnstahl:1992pi,Jin:1993up,Jeong:2012pa,Ohtani:2016pyk,Cai:2019vsg}
\begin{align}
\Pi_p^e &=    -\frac{1}{64\pi^4}s^2\ln(-s)
    + \frac{1}{9\pi^2}(4\langle \theta_u\rangle + \langle\theta_d\rangle)
    \nonumber\\ & \qquad
    -\frac{1}{32\pi^2}\ln(-s)\langle G^2\rangle+
    \frac{1}{36\pi^2}\ln(-s)\langle \theta_g\rangle
    \nonumber\\&\qquad
    -\frac{2}{3}\langle\bar uu\rangle^2
    -\frac{4}{3}\langle u^\dag u\rangle^2
     +r.t. \label{eq.Pi^e(rhoB)}
    \\
\Pi_s^e &=\frac{1}{4\pi^2}\ln(-s)\langle \bar dd\rangle 
    -\frac{1}{12}\langle G^2\rangle\langle\bar dd\rangle
    +r.t.,
\end{align}
where 
\begin{equation}
    \langle \theta_q\rangle \equiv \left\langle \bar q\left[ i\gamma_0 D_0  -\frac{m_q}{4}\right]q\right\rangle
\end{equation} 
is the zero component of the energy momentum tensor $T_{00}^q$ associated to quarks.
Similarly, 
\begin{equation}
    \langle \theta_g\rangle \equiv \left\langle \frac{\alpha_s}{\pi}\left[ G^a_{0\alpha}{G^{a}_0}{^\alpha}-\frac{1}{4}G^a_{\alpha\beta}G^{a\, \alpha\beta}\right]\right\rangle.
\end{equation}
At the end of each form factor, $r.t.$ stands for {\it regular terms}, i.e. those without branch cuts or poles in the complex $s$ plane, having no contribution to the FESR.

The condensates and nucleon mass as a function of baryon density can be described as 
\begin{align}
 \langle q^\dag q\rangle &= \frac{3}{2}\rho_B\\
  \langle\bar qq\rangle &= \langle\bar qq\rangle_0 [1-0.329(\rho_B/\rho_0)] \\
\langle G^2\rangle &=\langle G^2\rangle_0[1-0.066(\rho_B/\rho_0)]\\
\langle \Theta_q\rangle &=2.6\times 10^{-4}(\rho_B/\rho_0)\\
\langle \Theta_g\rangle &= 6.1\times 10^{-5}(\rho_B/\rho_0)\\
  m_N^* &=m_N[1-0.329(\rho_B/\rho_0)]
  \end{align}
where $\rho_B$ is the baryon density and $\rho_0=0.16 \mathrm{\, fm}^{-3} = 1.22\times 10^{-3}\mathrm{\, GeV}^3$ is the nuclear density.
In the estimations considered above, we have used the value of the pion-nucleon sigma term as $\sigma_N=45\textrm{ MeV}$.

\section{Magnetic field effects}

When considering finite magnetic field effects together with finite density, the required Feynman diagrams in the operator product expansion will involve a plethora of new condensates. 
This is  a consequence of Lorentz symmetry breaking. We have taken into account here a subset of all possible condensates. 
In order to have a simple representation we restrict the condensates to the relevant ones at finite density. 
The only strictly magnetic condensate to be considered is the polarization of the chiral condensate $\langle \bar q\sigma_{12} q\rangle$.
Hence,  we have the common terms that already exist in vacuum described in Eqs. (\ref{eq.Nuclear_FESR_p}) and (\ref{eq.Nuclear_FESR_s}), the exclusive density terms in Eq.\,(\ref{eq.Pi^e(rhoB)})\, the exclusive magnetic field terms \cite{Dominguez:2018njv}, and some combination of both that we need to determine.
Considering only the subset of condensates described above, the only relevant mixed contribution will be ${\Pi_u^e}\sim (eB)^2\langle q^\dag q\rangle$, which contributes in the hadronic sector to $\Delta\mu$ defined in Eq.\,(\ref{eq.Pi_N-had}), and  which is not considered in this work.

The set of FESR then for protons is the following:
 \begin{align}
 \lambda_p^2 &=  \frac{s_p^3}{192\pi^4}  
    + \frac{s_p}{32\pi^3}\langle\alpha_s G^2\rangle 
    + \frac{2}{3}\langle\bar uu\rangle^2 
    \nonumber\\&\quad
        +\frac{s_p}{2\pi^4}e_u e_d B^2
    +\frac{ s_p}{6\pi^4}(e_uB)^2[\ln(s_p/8m_q^2)-1]
   \nonumber\\&\quad  
    +\frac{s_p}{96\pi^4}(e_dB)^2\left[8\ln(s_p/8m_q^2)-9\right]
    \nonumber\\&\quad
    +3\rho_B^2
  - \frac{s_p}{9\pi^2} \langle \theta_d\rangle
  -\frac{4s_p}{9\pi^2}\langle \theta_u\rangle
  -\frac{s_p}{72\pi^2}\langle \theta_g\rangle
  \label{eq.lambda_p-B,rho1}
  \\
  &\nonumber\\
  \lambda_p^2 m_p  &= -\frac{s_p^2}{8\pi^2}\langle\bar dd\rangle
    +\frac{1}{12\pi}\langle \alpha_sG^2\rangle\langle\bar dd\rangle
        \nonumber\\&\quad
    +\frac{s_p}{2\pi^2}e_uB\langle \bar d\sigma_{12}d\rangle
    \nonumber\\&\quad
    +\frac{4}{3\pi^2}(e_u B)^2 \left[\ln(s_p/m_q^2)-1\right]\langle\bar dd\rangle
    \label{eq.lambda_p-B,rho2}\\
  &\nonumber\\
  -\lambda_p^2 \frac{\kappa_p B}{2} &= \frac{s_p^2}{48\pi^2}\langle \bar d\sigma_{12}d\rangle
   + \frac{e_u B\,s_p}{24\pi^2}\langle\bar dd\rangle \label{eq.lambda_p-B,rho3}
  \end{align}

  For neutrons, the FESR are the same with the replacement of $p\to n$ and $u\leftrightarrow d$.
 Here, $e_u$ and $e_d$ are the quark charges, $s_p$ is the continuum threshold for protons, and $\kappa_p$ is the proton anomalous magnetic moment.
 Note that the exclusive finite density contribution only appears in the last line of Eq.\,(\ref{eq.lambda_p-B,rho1}).
 
The next step is to incorporate density and magnetic field depending condensates. 
Besides the chiral condensate, there is not much in the literature for the value of the different condensates including both effects. 
On the other hand, most of the works which consider magnetic evolution of condensates usually consider very high magnetic field strength values, much bigger than 0.1\,GeV$^2$, so this is not a simple task.
Therefore we need to assume some simplifications in order to evaluate the sum rules. 

The evolution of the gluon condensate with respect to the magnetic field is almost negligible for low $B$ \cite{DElia:2015eey,Dominguez:2018njv}.
In this way, it is reasonable to consider only a baryon density dependent gluon condensate. 
From \cite{Hutauruk:2021dgv} we can see that, at low $B$, the approximation $f(B,\rho_B)\approx f(B,0)f(0,\rho_B)/f(0,0)$ can be applied to the quark condensate. 
We will use this approximation also for the nucleon mass, with the magnetic evolution of chiral condensate and nucleon mass the same used in \cite{Dominguez:2020sdf}.
In the case of the expectation values $\langle \theta_q\rangle $ and $\langle \theta_g\rangle $, we will leave them as baryon density dependent only.

\section{Results}

\begin{figure}
\includegraphics[scale=0.4]{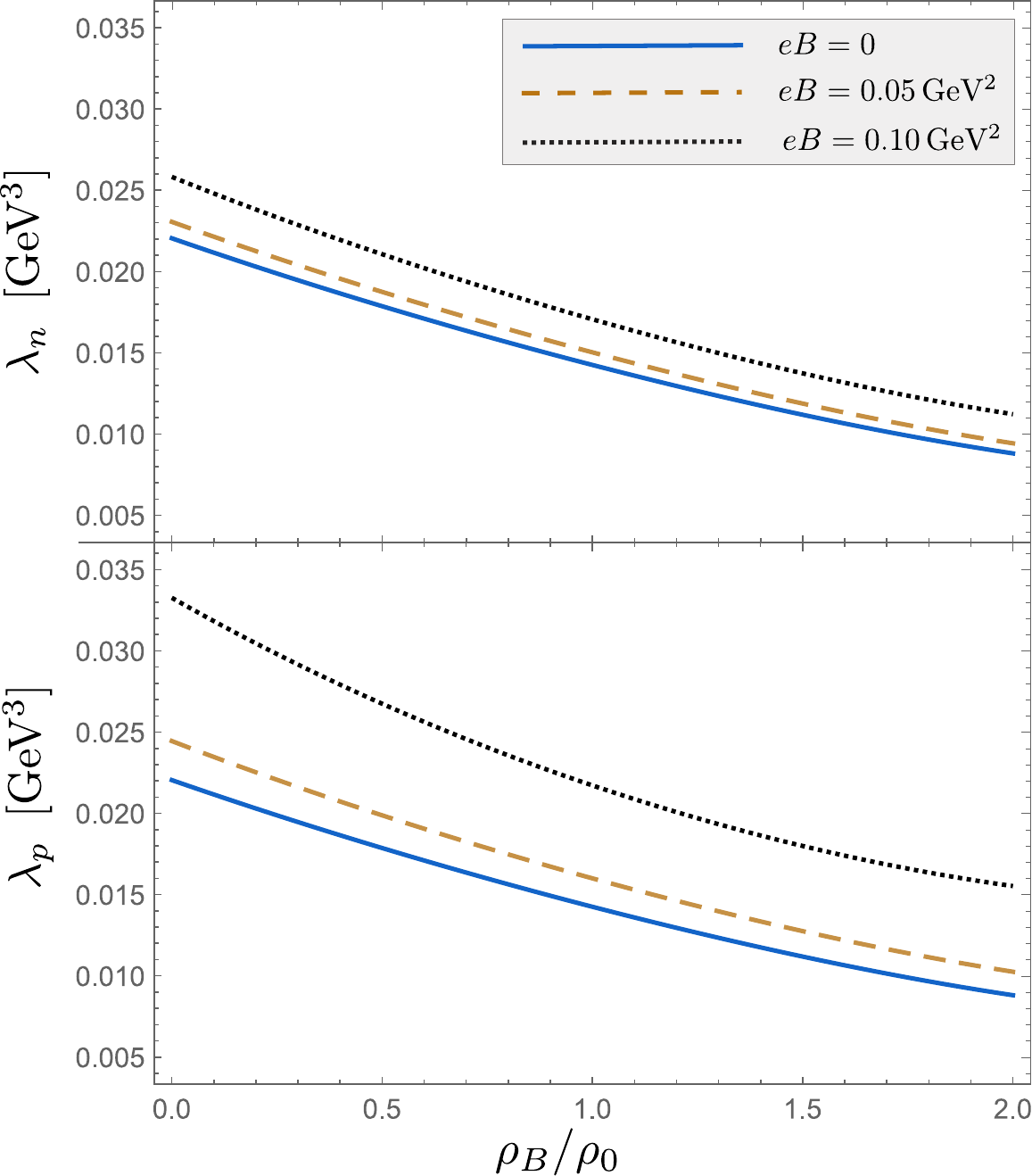}    
\caption{Neutron-current coupling (upper panel) and proton-current coupling (lower panel) as function of the baryon density in units of the nuclear density for different values of the external magnetic field strength.}
\label{fig.lambda_rho}
\end{figure}

\begin{figure}
\includegraphics[scale=0.44]{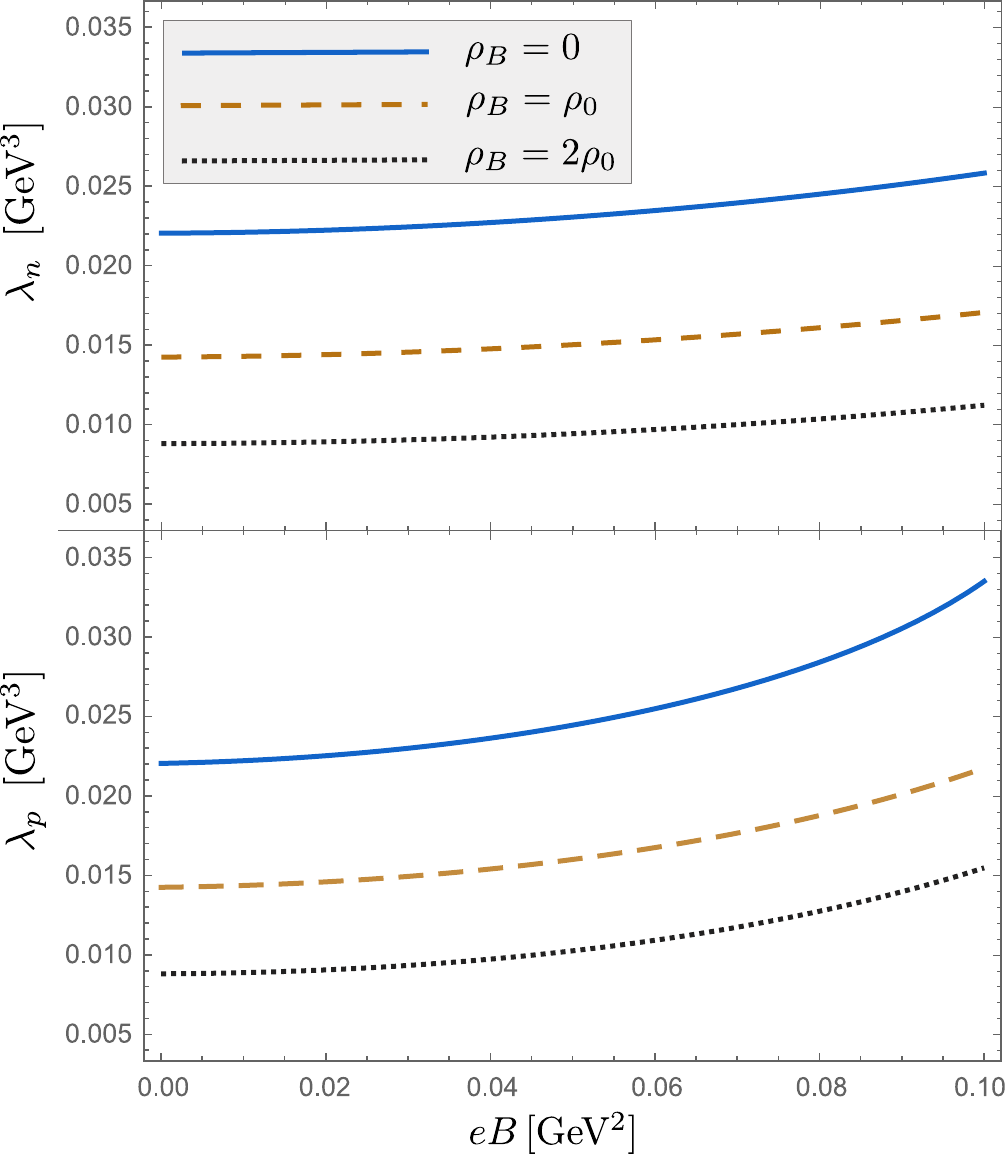}    
\caption{Neutron-current coupling (upper panel) and proton-current coupling (lower panel) as functions of the the external magnetic field strength for different values of baryon density in units of the nuclear density.}
\label{fig.lambda_eB}
\end{figure}

Once we have solved the set of the three coupled FESR described in Eqs. (\ref{eq.lambda_p-B,rho1})-(\ref{eq.lambda_p-B,rho3}), we obtain the magnetic and baryon-density evolution of the hadronic thresholds and the current couplings.
In Fig\,\ref{fig.lambda_rho} we show the behavior of the neutron and proton current couplings as a function of the baryon density, measured in units of the nuclear density $\rho_0 = 0.16\text{ fm}^{-3}$, for different values of the external magnetic field. 
The couplings diminish as the baryon density increases, in agreement with the deconfining tendency associated to high baryon density. 
On the contrary, as was noted in \cite{Villavicencio:2022gbr}, the presence of the magnetic field increases the values of the nucleon couplings.
In a similar way, in Fig.\,\ref{fig.lambda_eB}, we see the nucleon couplings as a function of the magnetic field for different values of the baryon density. 
The tendency of the couplings is to increase with the magnetic field, expressing the confining effect of the magnetic field. 
Both effects, magnetic field and baryon density, compete with each other.

\begin{figure}
\includegraphics[scale=0.4]{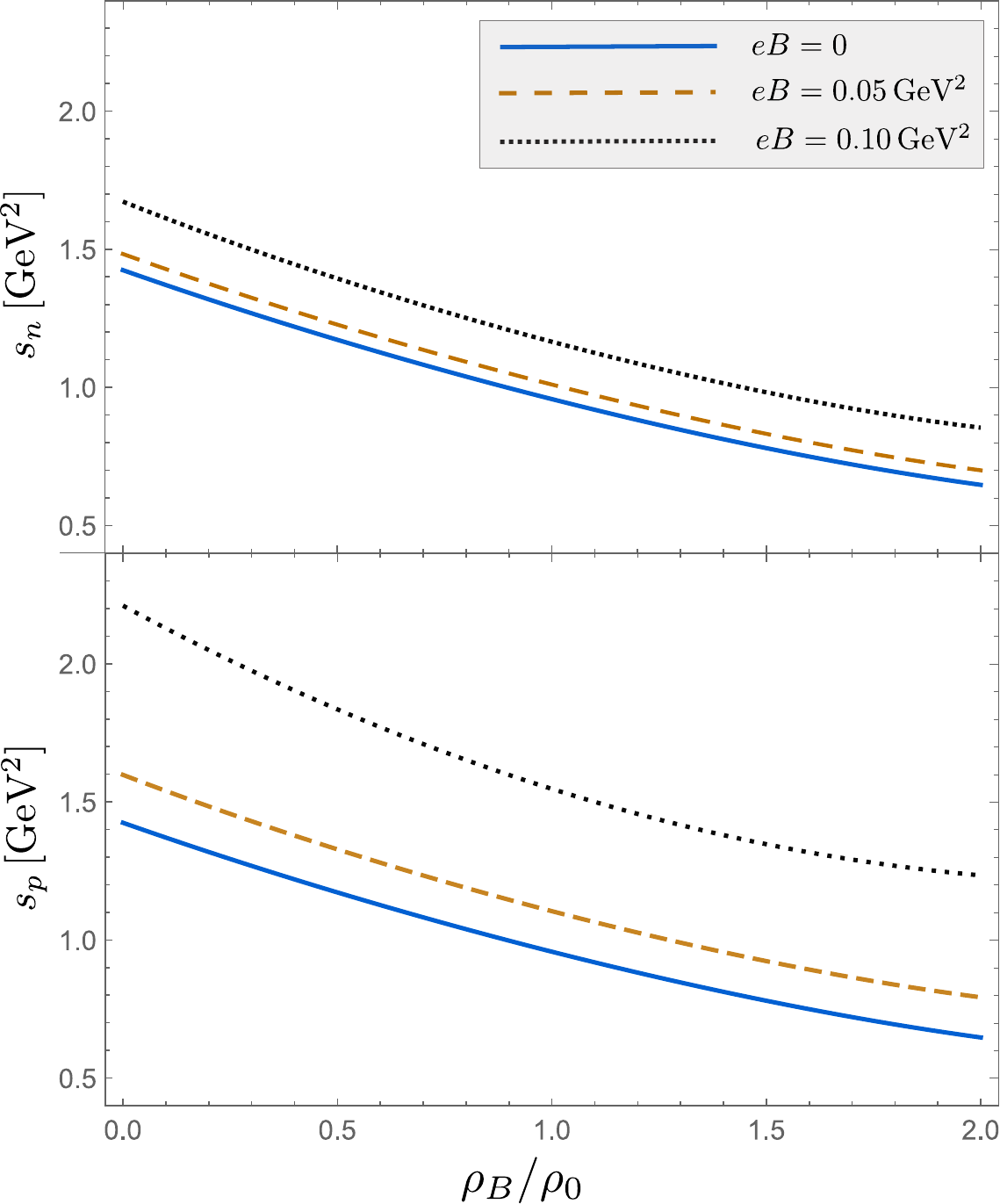}    
\caption{ Neutron-current continuum threshold (upper panel) and proton-current continuum threshold (lower panel) as functions of the baryon density in units of the nuclear density for different values of the external magnetic field strength.
}
\label{fig.s0_rho}
\end{figure}

\begin{figure}
\includegraphics[scale=0.43]{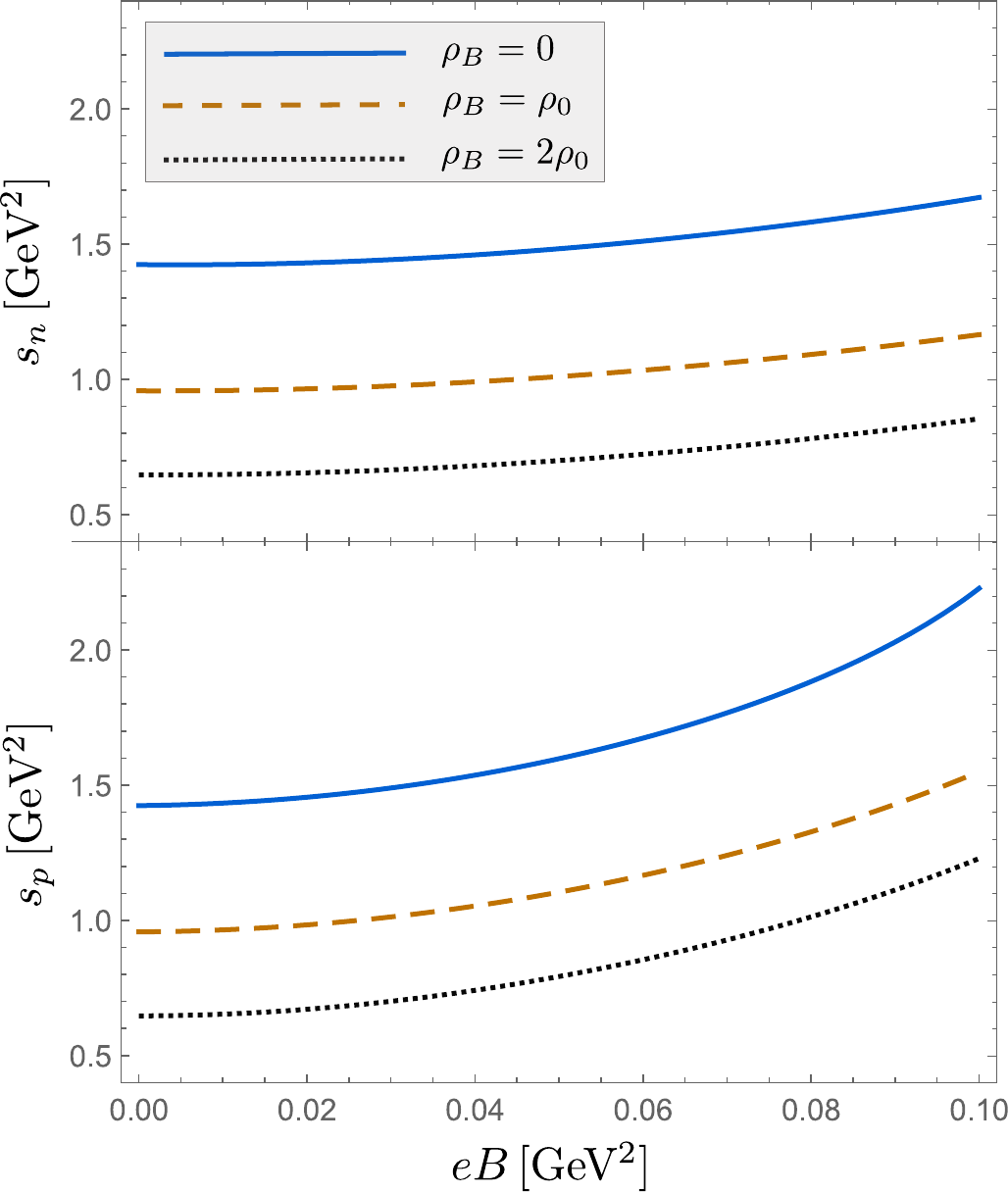}   
\caption{Neutron-current continuum  threshold (upper panel) and proton-current continuum threshold (lower panel) as functions of the the external magnetic field strength for different values of baryon density in units of the nuclear density.}
\label{fig.s0_eB}
\end{figure}

The hadronic thresholds for neutron and proton as a function of the baryon density for different values of the magnetic field are plotted in Fig.\,\ref{fig.s0_rho}. 
Notice the similarity of the shapes of the hadronic thresholds compared with the evolution of the nucleon couplings. 
This is somehow expected since, at lowest order in the OPE expansion, $\lambda_N^2\sim s_0^3$ as can be seen in Eq.\,(\ref{eq.lambda_p-B,rho1}). 
The decreasing of the hadronic thresholds as a function of the baryon density also is a signature of deconfinement trend.
The presence of the magnetic field raises the value of the thresholds.
The same can be seen in Fig.\,\ref{fig.s0_eB}, where the hadronic thresholds are plotted as a function of the magnetic field for different values of the baryon density. 

Notice that $s_n < s_p$ in all cases shown in Fig.\,\ref{fig.s0_rho} and Fig.\,\ref{fig.s0_eB}. 
This implies that, from Eq.\,(\ref{eq.gA(p,n)}), the axial-vector coupling constant at finite baryon density and magnetic field is then
\begin{equation}
    g_A=\frac{1}{48\pi^4}\frac{s_n^3}{\lambda_n\lambda_p}.
\end{equation}

\begin{figure}
\includegraphics[scale=0.40]{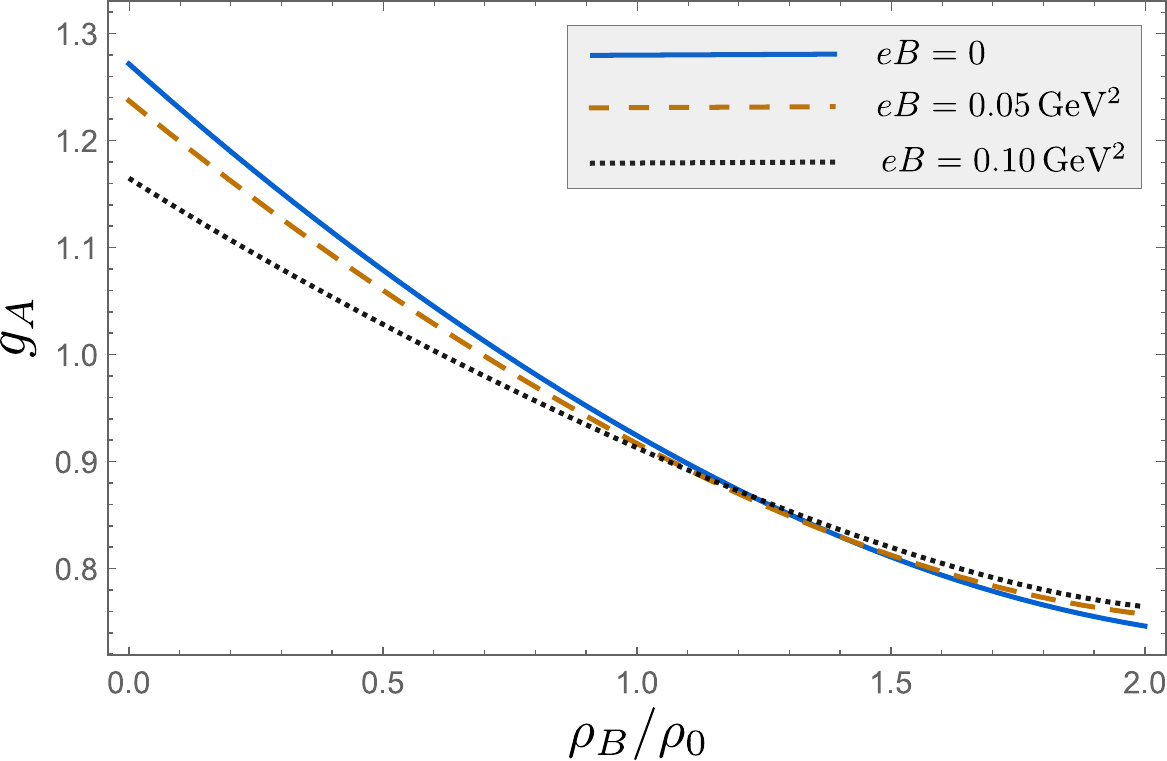}
\caption{Nuclear axial-vector coupling constant as functions of the baryon density in units of the nuclear density for different values of the external magnetic field strength.}
\label{fig.gA_rho}
\end{figure}

\begin{figure}
\includegraphics[scale=0.47]{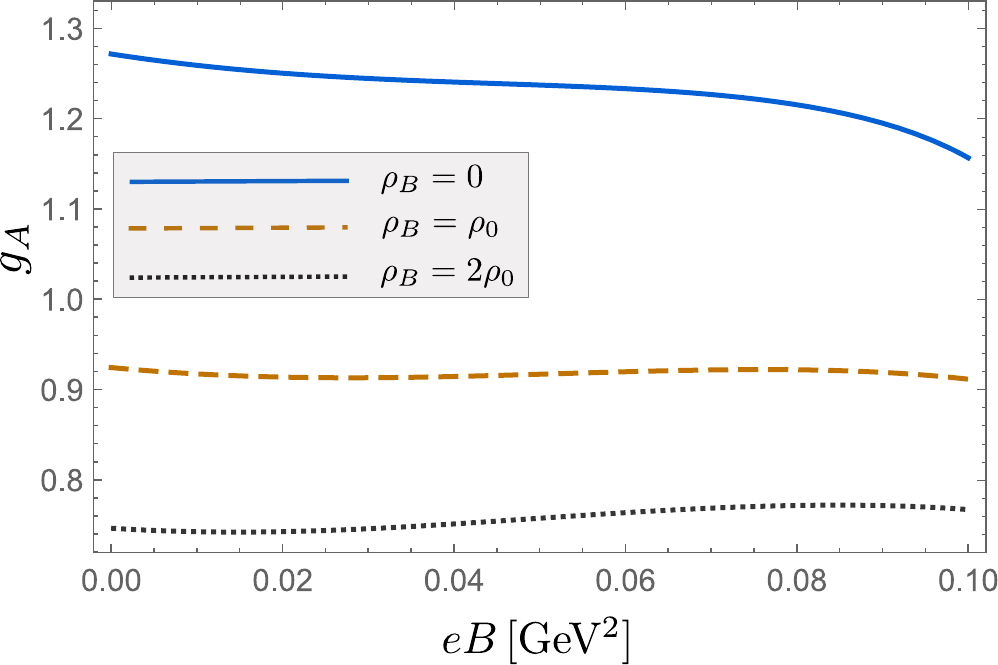}
\caption{Nuclear axial-vector coupling constant as a function of the the external magnetic field strength for different values of baryon density in units of the nuclear density.}
\label{fig.gA_eB}
\end{figure}

Since we have obtained the evolution of current couplings and neutron continuum threshold, the medium evolution of the nucleon axial-vector coupling constant is determined as can be seen in Fig.\,\ref{fig.gA_rho} as a function of the baryon density and in Fig.\,\ref{fig.gA_eB} as a function of the magnetic  field strength.
The axial-vector coupling decreases with the baryon density, and at the nuclear density $g_A^*\equiv g_A(\rho_0)   \approx 0.92$ in agreement with other approaches that suggest that $g_A^* \lesssim 1$.
As the magnetic field increases, the axial-vector coupling becomes smaller for $\rho_B\lesssim 1.25\rho_0$. 
However, for $\rho \gtrsim 1.25 \rho_0$, $g_A$ seems to increase with the magnetic field.
This means that in magnetar's nuclear matter, $g_A \sim g_A^*$  is apparently not affected by the magnetic field.

\section{Conclusions and discussion}

In this study, we conducted a comprehensive analysis within the framework of QCD FESR, investigating the evolution of the nuclear axial-vector coupling with respect to both baryonic density and a constant, uniform magnetic field. 
Our research aims to emulate the scenario of nuclear matter in magnetars.

As a result, we obtain a decreasing axial-vector coupling when baryonic density increases, where in the specific case of the nuclear density $g_A^* \approx 0.92$ in accordance with the commonly assumed value that establishes that $g_A^*\sim 1$ \cite{Rho:1974cx,Wilkinson:1974huj,Brown:1978zz,Park:1997vv,Carter:2001kw,Lu:2001mf,Suhonen:2017krv,Bass:2020bkl,Rho:2021zwm,Rho:2023vow}.
The axial-vector coupling at nuclear density is basically unchanged by the effects of the magnetic field, at least for the values considered here. 

The plotted curves of the axial-vector coupling as a function of the baryon density present an intersection when different values of the magnetic field are considered.
This behavior, the increasing of $g_A$ with the magnetic field, may be enhanced for higher values of baryon density and magnetic field, which is the case when one gets close to the core of the magnetar.

It is important to remark that there are many approximations concerning  the magnetic evolution of the different condensates. 
Since there are no other approaches which consider both, baryonic density and magnetic field effects, in the axial-vector coupling it is difficult to determine if this behavior is correct, but This simplifications is taken under reasonable basis. 
However it is important to obtain complete and better in-medium evolution of the condensates as well as the incorporation of other neglected condensates. 
This will be explored elsewhere in future work.

\begin{acknowledgments}
M.L., C.V. and R.Z. acknowledge support from ANID/CONICYT FONDECYT Regular (Chile) under Grants No. 1190192, No. 1200483 and No.1220035. M. L. acknowledges also support from ANID PIA/APOYO AFB 180002 (Chile) and from the Programa de Financiamiento Basal FB 210008 para Centros Cient\'{\i}ficos y Tecnol\'ogicos de excelencia de ANID
\end{acknowledgments}


%

\end{document}